\documentclass[a4paper,twocolumn,prl]{revtex4}%
\usepackage[colorlinks,linkcolor=blue,urlcolor=blue,citecolor=blue]{hyperref}

\usepackage[latin9]{inputenc}
\usepackage{amsmath}
\usepackage{amssymb}
\usepackage{graphicx}
\usepackage{comment}
\usepackage{mathrsfs}
\usepackage{amsfonts}
\usepackage{float}%
\usepackage{multirow}
\providecommand{\U}[1]{\protect\rule{.1in}{.1in}}
\makeatletter

\DeclareFontEncoding{LGR}{}{}
\DeclareTextSymbol{\~}{LGR}{126}
\@ifundefined{textcolor}{}
{
\definecolor{BLACK}{gray}{0}
\definecolor{WHITE}{gray}{1}
\definecolor{RED}{rgb}{1,0,0}
\definecolor{GREEN}{rgb}{0,1,0}
\definecolor{BLUE}{rgb}{0,0,1}
\definecolor{CYAN}{cmyk}{1,0,0,0}
\definecolor{MAGENTA}{cmyk}{0,1,0,0}
\definecolor{YELLOW}{cmyk}{0,0,1,0}
}

\makeatother
\begin{document}
%\title{Exceptional Flexoelectric Effect in Silicon Thin Film under Bending}
%\title{A New Mechanism of Flexoelectricity Beyond the Linear Response Theory}
%\title{Exceptional Flexoelectric Effect of Silicon: A New Mechanism of Flexoelectricity}
\title{ Giant Flexoelectricity in Bent Silicon Thinfilms}
%\title{Mechanism of Giant Flexoelectricity in Silicon Thinfilms Beyond the Linear Response Theory}
\author{Dong-Bo Zhang$^{1,4}$}
\thanks{Corresponding author}
\email[]{dbzhang@bnu.edu.cn}
\author{Kai Chang$^{2,3,5}$}
\thanks{Corresponding author}
\email[]{kchang@semi.ac.cn}
\affiliation{$^{1}$College of Nuclear Science and Technology, Beijing Normal University, Beijing 100875, P.R. China}
\affiliation{$^{2}$SKLSM, Institute of Semiconductors, Chinese Academy of Sciences, P.O. Box 912, Beijing 100083, China}
\affiliation{$^{3}$CAS Center for Excellence in Topological Quantum Computation, University of Chinese Academy of Sciences, Beijing 100190, China}
\affiliation{$^{4}$Beijing Computational Science Research Center, Beijing 100193, P.R. China}
\affiliation{$^{5}$Beijing Academy of Quantum Information Sciences, Beijing 100193, China}
\begin{abstract}
We reveal that strong flexoelectric effect of solids can be induced due to the significant charge migration along the strain gradient direction, which represents a new understanding of the origin of flexoelectricity.  Beyond the linear response theory, we illustrate such charge migration that is driven by an electric field effect in bent silicon thinfilms. Due to such charge migration, the variation of atomic charge no longer represents a linear response to strain gradient and the resulting giant flexoelectric coefficients being size dependent cannot be treated as a bulk property.  The obtained flexoelectric coefficients compare well with the typical experimental values as reported in various ceramics. Our results shed light on elucidating the discrepancy between theory and experiment, and pave a new way to discover excellent flexoelectric performance in conventional materials.
\end{abstract}

%\pacs{75.50.Dd, 71.70.Fk, 73.20.Pr, 72.25.Dc}
\maketitle
Recent experimental and theoretical investigations have substantially advanced our knowledge of flexoelectricity, a phenomenon that depicts the coupling between the electric polarization and the strain gradient in dielectrics~\cite{wang,Kogan}. Being a universal effect of any structure without symmetry limitation~\cite{review2}, flexoelectric effects have been identified in various systems such as  ceramics~\cite{Cross1,Cross2,Cross3,Sharma3}, hybrid-semiconductors~\cite{Catalan3,wang}, elemental crystals~\cite{pv,Hong3}, and even soft materials~\cite{polymer1,polymer2}. However, our understanding of flexoelectricity is not complete yet. For example, for the flexoelectric coefficient, the significant discrepancy between calculations and experiments still remains elusive~\cite{review1,review2,review3,Tagantsev3}. Flexoelectricity is important for both fundamental research and developing applications. It not only can be a substitute for piezoelectricity in many occasions~\cite{review2,Catalan1,Cross5,Catalan5,Sharma2}, but also may enable practical applications~\cite{pv,Catalan4} where piezoelectricity does not approach.

Microscopically, flexoelectricity contains both electronic and lattice-mediated~\cite{review3} effects, which have been delineated with the linear response theory~\cite{martin,Tagantsev5,Resta1}. The essence of this theory is to establish the response function of charge density~\cite{Resta1} and atomic displacement~\cite{Tagantsev,Sharma1} with respect to strain gradient. However, although the experiments have revealed high flexoelectric coefficients in several bulk solids to be at the level of $10^{-6}$C/m or even higher, the prediction based on the linear response theory is orders of magnitude lower than the experimental data~\cite{Cross1,Cross2,Cross6,Sharma3}. Such striking underestimation brings a natural question: Can certain mechanism of flexoelectricity exist out of the regime of the linear response theory? Theoretically, atomistic simulations are the method of choice to address this issue.  Unfortunately, efforts in this aspect are still missing, partially due to the difficulty of simulating flexoelectric systems subject to realistic inhomogeneous deformations such as bending with standard methods~\cite{vasp,qe}.

Here, based on the generalized Bloch theorem~\cite{gbt1,gbt2,gbt3}, we reveal strong flexoelectric effects in crystalline silicon through a new mechanism where electronic charges migrate along the strain gradient. In this way, electric polarizations form also in the strain gradient direction, giving rise to unexpectedly high flexoelectric coefficients. We use (100) silicon films with various thickness~\cite{film1,film2} to demonstrate this idea. Our quantum mechanical simulations of the bent silicon films illustrate the charge migration from the compressive side to the tensile side, which is driven by a strain-induced electric field. The explanation is derived by analyzing the response of the electronic states to mechanical bending. What is interesting is that because that the variation of the atomic charge is  dominated by the local strain, it does not represent a linear response to strain gradient.

With this, the dipole moment and the flexoelectric coefficient are next calculated. Surprisingly, the flexoelectric coefficient displays a square scaling with the film thickness and approaches  the level of $10^{-6}$C/m at 75 nm thickness, being comparable with the typical experimental values as reported in various ceramics~\cite{Cross1,Cross2,Cross6,Sharma3}. Its size dependent behavior also hints that although the flexoelectric effect mainly origins from the bulk contribution, the flexoelectric coefficient can not be treated as a bulk property of material.

A stress-free (100) silicon film can be depicted by a square lattice with a translation periodicity of $T_0$, where the unit cell contains a relatively small number of atoms, $N$. The precise value of $T_0$ can be determined by first-principles and density-functional tight-binding (DFTB)~\cite{dftb1,dftb2} calculations. For example, a (100) silicon film with a 10 nm thickness, DFTB calculation gives $T_0=5.43~\text{\AA}$. However, a bending deformation breaks the translational symmetry along the principal curvature. This makes first-principles and other QM calculations formulated with periodic boundary conditions intractable. Instead, we employ the generalized Bloch scheme~\cite{gbt1,gbt2,gbt3} implemented into the DFTB method. In this scheme, the bent film is described with basic repetition rules of translation ${\bf T}$ and rotation of angle $\theta$ performed in the curvilinear coordinate, Fig.~1(a),
\begin{equation}
\label{bending}
{\bf X}_{\xi,\lambda,n}=\xi{\bf T}+{\bf R}^{\lambda}(\theta){\bf X}_{n},
\end{equation}
where, ${\bf X}_{n}$ represents atoms inside the primitive repeating cell and ${\bf X}_{\xi,\lambda,n}$ represents the atoms inside the replica of the repeating cell indexed by ($\xi,\lambda$).  The repeating cell contains the same $N$ atoms in the unit cell of the undistorted film. For example, for the (100) silicon film with 10~nm thickness, $N=162$. Index $n$ runs over the $N$ atoms inside the cell. The relatively small $N$ allows for systematic QM simulations of the bent structure. We have carried out DFTB calculations on a series of bent (100) silicon films with thickness $(h)$: $10<h<200$~nm. The exposed surfaces of the films are hydrogenated. The atomic structures are optimized via a conjugate gradient energy minimization to the repeating cell. More computational details are provided in the Supplemental material.

\begin{figure}[tb]
\includegraphics[width=1\columnwidth]{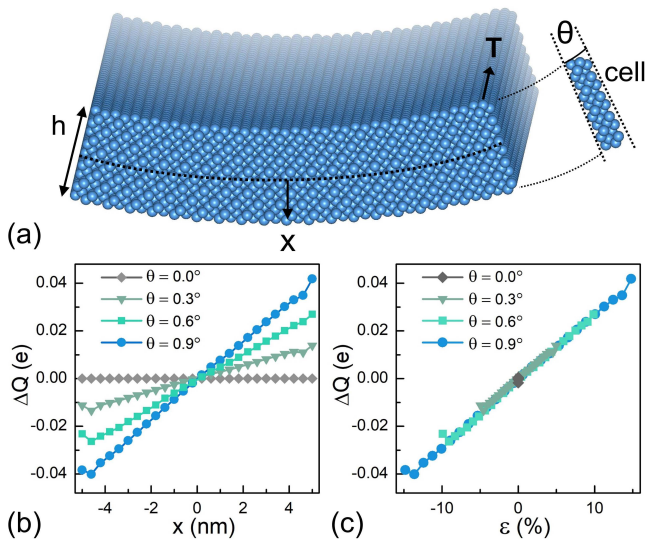}
\caption{(a) A bent (100) silicon film with (100) symmetry planes at the surfaces (left) and the primitive repeating cell (right). ${\bf T}$ is the translation vector that is invariant against bending. $h$ denotes the film thickness and $-h/2<x<h/2$ along the thickness dimension measures the distance from the neutral surface of the bent film. The bending angle corresponding the repeating cell is $\theta$. (b) Distribution of neat charge per atom $\triangle Q$ along the film thickness dimension, $x$, of the 10~nm thick bent film at different bending angles $\theta$. (c) $\triangle Q$ versus the local strain $\varepsilon$  of the 10~nm thick bent film. Note that $\theta=0.0^{\circ}$ corresponds to the stress-free film.}%
\label{field}%
\end{figure}

We first calculate the variation of the atomic charge in the bent film,
\begin{equation}
\label{charge}
\triangle Q=Q_{\text{bent}}(x)-Q_{\text{stress-free}}(x),
\end{equation}
where, $Q_{\text{bent}}$  refers the electronic charge of the atom located at position $x$  of the bent film and $Q_{\text{stress-free}}$ is the same but for the stress-free film. Note that the nuclear charges of individual atoms that remain constant against bending are not considered. The atomic electronic charge, $Q$, is obtained via a Mulliken charge analysis (see the Supplemental material for details). Fig.~1(b) displays the obtained $\triangle Q$ of the 10 nm thick silicon film. At each bending angle $\theta$, $\triangle Q$ varies almost linearly along the thickness dimension, $x$, except those atoms belonging to the film surface. What is important is that on the tensile side of the bent film ($x<0$), $\triangle Q>0$ indicates that atoms in this region gain electronic charges. On the contrary, on the compressive side ($x>0$), $\triangle Q<0$ indicates that atoms in this region lose electronic charges. Because the film as a whole is charge neutral, this result reveals a charge transfer from the compressive side to tensile side of the bent film.

We also indicate that the amount of $\triangle Q$ is related to the local strain ($\varepsilon$). In a bent film, the inhomogeneous strain field is simply $\varepsilon(x)=x/\tau=(\theta/T_0) x$, with $1/\tau=\theta/T_0$ being the curvature of the neutral surface, Fig. 1(a).  With this relation, we find that data obtained under different bending angles, $\theta$, collapses when plotting $\triangle Q$ versus $\varepsilon$, Fig.~1(c). Such scaling collapse hints that the $\triangle Q$ response  is solely dominated by the  local strain in the bent film and allows us to approximate that,
 \begin{equation}
\label{g}
\triangle Q\simeq \chi\varepsilon(x)=\chi gx,
\end{equation}
ignoring the atoms on the surface. Here, $\chi$ denotes the slope and $g=\partial\varepsilon/\partial x=\theta/T_0$ is the strain gradient along the film thickness dimension. This revelation is contrary to the prediction of the linear response theory where the charge variation induced by strain gradient distributes homogeneously throughout the deformed solids~\cite{Tagantsev5,Resta1}. Further, the distribution of $\triangle Q$, Fig.~1(b), also hints that the charge accumulation on the film surfaces due to bending is not significant.

\begin{figure}[tb]
\includegraphics[width=1\columnwidth]{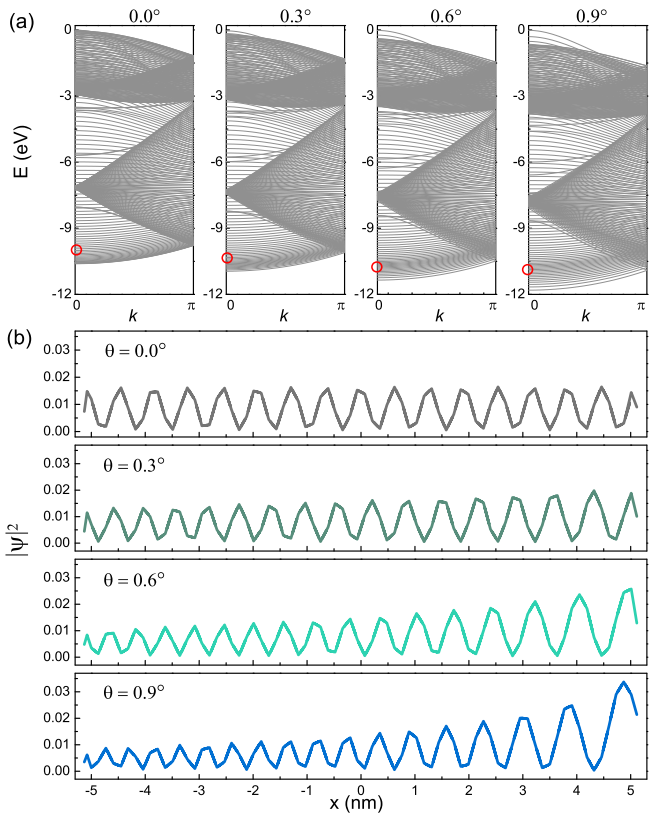}
\caption{(a) Comparison of electronic band structures of occupied states of stress-free film ($\theta=0.0^{\circ}$) and bent (100) silicon films ($\theta=0.3^{\circ}, 0.6^{\circ},0.9^{\circ}$). The Fermi energy is set at zero. (b) Wave function spatial distributions ($|\Psi|^2$) along the film thickness dimension, $x$, for the electronic states whose locations in the energy spectrum are indicated by open circles in (a) at $k=0$.  }%
\label{bond}%
\end{figure}

Insight into the charge transfer in the bent film can be obtained by analyzing the impact of bending on the electronic structure. First of all, we focus on the variation of the spatial distribution of electronic states. Fig.~2(a) displays the band structure of occupied states with wavenumber $k$ in [001] direction for the stress-free (100) silicon film ($\theta=0.0^{\circ}$), and accordingly, the band structures with $k$ at $l=0$ for the bent film under different bending angles, $\theta$. Note that in the generalized Bloch theorem, electronic states are indexed by both the continuous wavenumber $k$, and the discrete rotational quantum number, $l$ (see the Supplemental material for more details). Using the state marked by the open circle in Fig.~2(a) as an example, Fig.~2(b) displays the distribution of its wave function along the film thickness dimension for both the stress-free film and the bent film. We note that electronic states around the open circle are non-degenerate. Thus,  for a given state in the stress-free film, it is possible to identify the corresponding state in the bent film by simply referring its order in the energy spectrum. For the stress-free film, the wave function  essentially adopts a uniform distribution. However, under bending, the amplitude of the wave function increases on the tensile side while decreases on the compressive side. This result hints a charge immigration from the compressive side to the tensile side of the bent film, revealing a pronounced built-in electric field effect~\cite{electric}. The trend is more pronounced for the larger bending angles. More analysis on the state evolution with respect to bending is shown in the Supplemental material.

\begin{figure}[tb]
\includegraphics[width=1\columnwidth]{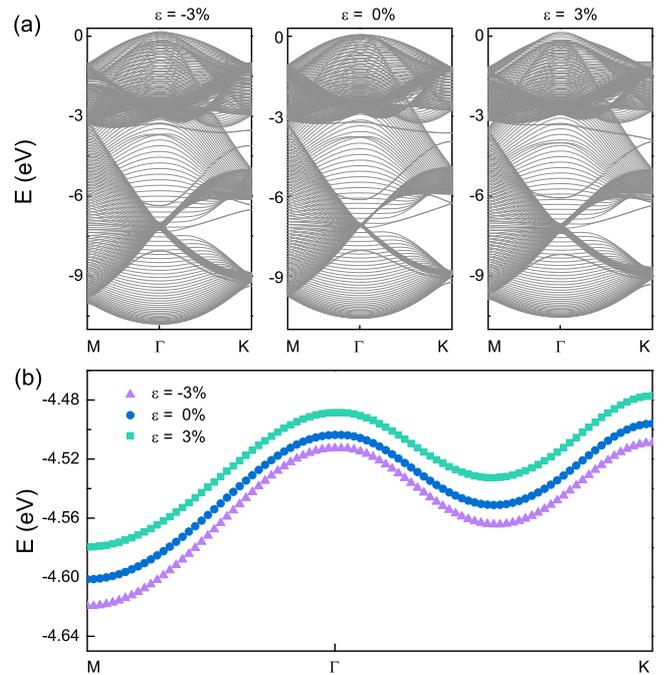}
\caption{ (a) Comparison of electronic band structures of occupied states of the (100) silicon film under different normal strains. (b) $E_{\text{ave}}$ of the (100) silicon film under different normal strains.  }%
\label{field}%
\end{figure}

Bending also has strong influence on the electronic energy spectrum, Fig.~2(a). To a great extent, this can be well understood by examining the electronic response of the stress-free film to normal strains. Considering tension and compression applied along the film thickness dimension, Fig.~3(a) displays the band structures of the strained 10 nm thick silicon film. Note that to illustrate the shift of the energy levels due to strain, the stationary vacuum energy level~\cite{wei1,wei2,Resta4} is chosen as a reference for the energy levels of the film electronic states under different strains. We find that only the compressed film adopts a significant increase in bandwidth. This indicates that the enlarged bandwidth of the bent film is essentially due to the compressional strain. Further, it is instructive to study the evolution of the averaged electronic energy of the film under normal strain,
 \begin{equation}
\label{energy}
E_{\text{ave}}(k)=\sum_\nu E_{\nu}(k),
\end{equation}
where, $E_{\nu}$ denotes the energy level of state $\nu$ with wavenumber $k$ of the film. The summation runs over all the occupied states. Fig.~3(b) shows that $E_{\text{ave}}$ adopts a downward (upward) shift for the film under compression (tension) for wavenumbers $k$ throughout the whole Brillouin zone. With this statistical result, we can infer that in the bent film, the electronic states on the tensile side will also adopt an upward shift in energy and those states on the compressive side do the reverse. This also represents a sign of the electric field effect.

\begin{figure}[tb]
\includegraphics[width=1\columnwidth]{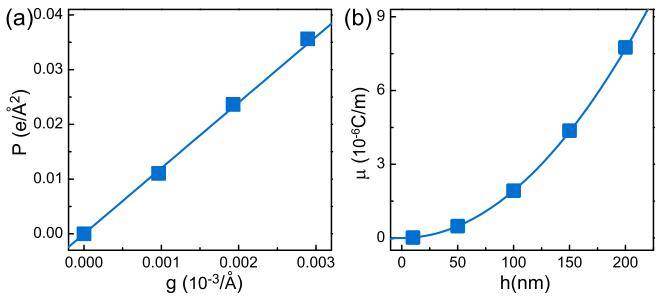}
\caption{(a) Flexoelectric polarization versus strain gradient of the bent 10~nm thick film. (b) Flexoelectric coefficient as a function of film thickness }%
\label{field}%
\end{figure}

We now estimate the bending-induced polarization ${\bf P}$, the core physical quantity to probe  flexoelectricity. Classically,  ${\bf P}$ can be obtained directly with the atomic charge. We set the coordinate origin at $x=0$. By symmetry, it is obvious that ${\bf P}\simeq0$ for the stress-free film. In this way, the net polarization along the film thickness dimension for the bent film can be obtained using Eq.~(\ref{charge}) as,
\begin{equation} \label{polar}
{\bf P}=\frac{1}{\Omega}\int_{-h/2}^{h/2}\triangle Q(x)xdx=\frac{\chi h^2}{12T_0^2}g,
\end{equation}
where, $\Omega=T_0^{2}h$ denotes the volume of the repeating cell.  Because for a given film, $T_0$, $h$, and even $\chi$ are all constant, $\bf P$ is proportional to the strain gradient $g$. As such, the flexoelectric coefficient
\begin{equation}\label{mu}
\mu={\bf P}/g=\chi h^2/12T_0^2,
\end{equation}
displays a square dependence on the film thickness, $h$.

However, calculating ${\bf P}$ directly from the atomic charge, Eq.~(\ref{polar}) completely ignores  the signature of the spatial distribution of electronic charge density. Instead, we employ a more rigorous approach where the electronic charge density ($\rho$) is explicitly accounted for~\cite{Resta3,Resta2,polar1},
\begin{equation} \label{realp}
{\bf P}=\frac{1}{\Omega}\int_{\Omega}\rho({\bf r}){\bf r}d{\bf r}.
\end{equation}
For a bent film, $\rho$ is provided by the generalized Bloch theorem (see the Supplemental material for details). In general, the integral should be over the whole space. However, because the charge transfer is along the film thickness dimension $x$, the induced ${\bf P}$ is essentially along $x$ as well. In this way, considering the rotational symmetry of the bent film, Eq.~(\ref{bending}), the integral can be conducted merely over the repeating cell as shown in Fig.~1(a).  It is worth to note that the surface effect~\cite{Tagantsev2,Stengel1,surface} is not important here, see the Supplemental material for more analysis. The origin of ${\bf r}$ is also set at $x=0$.

In practice, the integral in Eq.~(\ref{realp}) is evaluated numerically. We have calculated ${\bf P}$ of the bent 10 nm thick silicon film at different bending angles. Results shown in Fig.~4(a) reveals that ${\bf P}$ follows a simple linear dependence on strain gradient $g$. As such, the flexoelectric coefficient can be identified directly from the slope, i.e., $\mu={\bf P}/g$.
For the 10 nm thick silicon film, $\mu=1.9\times10^{-8}$~C/m. We have also carried out calculations of a series of (100) silicon films with thickness $h=50$, 100, 150, and 200~nm. For these films, ${\bf P}$ maintains the linear dependence on $g$, Fig. S3 of the Supplemental material. The $\mu$ values obtained using Eq.~(\ref{mu}), ranging from $0.5\times10^{-6}$ to $7.8\times10^{-6}$~C/m, exhibit a quadratic dependence on $h$ as guided by fitting the atomistic data as $\mu=C_0(h/\text{nm})^2$ with $C_0=1.9\times10^{-10}$~C/m, Fig.~4(b). This relation corroborates our finding with Eq.~(\ref{mu}), and allows us to figure out that $\mu$ is already at the level of $10^{-6}$~C/m when $h\geq75$~nm. Note that this $\mu$ is comparable with the experimental values of $\mu$ as found in various ferroelectric ceramics~\cite{Cross2,Cross4,Cross6,Cross7,Sharma3}.

The ultrahigh values of $\mu$ illustrate the importance of the new mechanism, and the above results reveal several important aspects that distinguish the new mechanism from the linear response theory.
(i) The variation of atomic charge $\triangle Q$ is dominated by the level of the local strain of the bent film and thus distributes inhomogeneously along the strain gradient direction. In this way, the induced charge ($\triangle Q$)  can not be treated as a well-defined linear response to strain gradient. Consequently, the obtained flexoelectric coefficients are  size dependent, no longer a bulk property of material.
(ii)According to the linear response theory, silicon is a system with weak flexoelectricity~\cite{Tagantsev3,Hong3}. However, the flexoelectric coefficient obtained here is exceptionally high, hinting that silicon may be an excellent candidate for flexoelectric applications. More fundamentally, this result also sheds light on understanding the long-standing discrepancy between theory and experiment~\cite{review1,review2,review3}.
(iii) As showcased with silicon, exploring the new mechanism relies on the capability to simulate realistically inhomogeneous deformation of solids, which is beyond the reach of atomistic approaches including first-principles calculations. This feature herein highlights the importance of the employed generalized Bloch scheme~\cite{gbt1,gbt2,gbt3}.

In summary, using the generalized Bloch theorem coupled with DFTB method and taking silicon films as an example, we reveal a new mechanism that can lead to strong flexoelectric effects in solids. With thorough electronic structure analysis of the bent silicon film, this new mechanism is attributed to a bending-induced electric field effect. This new mechanism cannot be interpreted by the linear response theory where a well-defined linear response of electrons  to the strain gradient is needed. Indeed, for silicon, the obtained flexoelectric coefficients are much larger than what is predicted by the linear response theory, but comparable with the typical experimental values as reported in ceramics. This suggests that silicon may be an excellent flexoelectric system, contrary to the conventional wisdom. As such, the new mechanism presents dual importance for the current flexoelectric studies. First, it may be critical to deepening our understanding on the flexoelectricity in solids. Second, due to it, a class of materials with excellent flexoelectric performance are likely to be discovered in conventional semiconductors, e.g., Ge, GaAs, etc.

This work was supported by  the MOST of China (Grants No. 2016YFE0110000 and No. 2017YFA0303400) and  NSFC under Grants Nos. 11674022, 11874088  and U1930402. D.-B.Z. was supported by the Fundamental Research Funds for the Central Universities. Computations were performed at the Beijing Computational Science Research Center and Beijing Normal University.


\begin{thebibliography}{99}
\bibitem {wang} L. Wang, S. L. Liu, X. Feng, C. Zhang, L. Zhu, J. Zhai, Y. Qin, and Z. L. Wang, Flexoelectronics of centrosymmetric semiconductors, {\it Nature Nanotechnol.} 10.1038/s41565-020-0700-y (2020).
\bibitem {Kogan} S. M. Kogan, Piezoelectric effect during inhomogeneous deformation and acoustic scattering of carriers in crystals, {\it Sov. Phys. Solid State} {\bf 5}, 2069 (1964).
\bibitem {review2} P. Zubko, G. Catalan, and A. K. Tagantsev, Flexoelectric effect in solids, {\it Annu. Rev. Mater. Res.} {\bf 43}, 387 (2013).
\bibitem {Cross1} L. E. Cross, Flexoelectric effects: Charge separation in insulating solids subjected to elastic strain gradients, {\it J. Mater. Sci.} {\bf 41}, 53 (2006).
\bibitem {Cross2} W. Ma and L. E. Cross, Large  flexoelectric polarization in ceramic lead magnesium niobate, {\it Appl. Phys. Lett.} {\bf 79}, 4420 (2001).
\bibitem {Cross3} W. Ma and L. E. Cross, Flexoelectric polarization of barium strontium titanate in the paraelectric state, {\it Appl. Phys. Lett.} {\bf 81}, 3440 (2002).
\bibitem {Sharma3} M. Gharbi, Z. H. Sun, P. Sharma, K. White, and S. El- Borgi, Flexoelectric properties of ferroelectrics and the nanoindentation size-effect, {\it Int. J. Solids Struct.} {\bf 48}, 249 (2011).
\bibitem {Catalan3} J. Narvaez, F. Vasquez-Sancho, and G. Catalan, Enhanced flexoelectric-like response in oxide semiconductors, {\it Nature} {\bf 538}, 219 (2016).
\bibitem {pv} M.-M. Yang, D. J. Kim, and M. Alexe, Flexophotovoltaic effect, {\it Science} {\bf 360}, 904 (2018).
\bibitem {Hong3} J. Hong and D. Vanderbilt, First-principles theory and calculation of  flexoelectricity, {\it Phys. Rev. B} {\bf 88}, 174107 (2013).
\bibitem {polymer1} X. Wen, D. Li, K. Tan, Q. Deng, and S. Shen, Flexoelectret: An electret with a tunable  flexoelectriclike response, {\it Phys. Rev. Lett.} {\bf 122}, 148001 (2019).
\bibitem {polymer2} B. Chu and D. R. Salem, Flexoelectricity in several thermoplastic and thermosetting polymers, {\it Appl. Phys. Lett.} {\bf 101}, 103905 (2012).
\bibitem {review1} P. V. Yudin and A. K. Tagantsev, Fundamentals of flexoelectricity in solids, {\it Nanotechnology} {\bf 24}, 432001 (2013).
\bibitem {review3} B. Wang, Y. Gu, S. Zhang, and L.-Q. Chen, Flexoelectricity in solids: Progress, challenges, and perspectives, {\it Prog. Mater. Sci.} {\bf 106}, 100570 (2019).
\bibitem {Tagantsev3} P. V. Yudin, R. Ahluwalia, and A. K. Tagantsev, Upper bounds for  flexoelectric coefficients in ferroelectrics, {\it Appl. Phys. Lett.} {\bf 104}, 082913 (2014).
\bibitem {Catalan1} G. Catalan, A. Lubk, A. H. G. Vlooswijk, E. Snoeck, C. Magen, A. Janssens, G. Rispens, G. Rijnders, D. H. A. Blank, and B. Noheda, Flexoelectric rotation of polarization in ferroelectric thin films, {\it Nature Mater.} {\bf 106}, 963 (2011).
\bibitem {Cross5} B. Chu, W. Zhu, N. Li, and L. E. Cross, Flexure mode flexoelectric piezoelectric composites, {\it J. Appl. Phys.} {\bf 106}, 104109 (2009).
\bibitem {Catalan5} U. K. Bhaskar, N. Banerjee, A. Abdollahi, Z. Wang, D. G. Schlom, G. Rijnders, and G. Catalan, A  flexoelectric microelectromechanical system on silicon, {\it Nature Nanotechnol.} {\bf 11}, 263 (2016).
\bibitem {Sharma2} M. S. Majdoub, P. Sharma, and T. Cagin, Enhanced size-dependent piezoelectricity and elasticity in nanostructures due to the  flexoelectric effect, {\it Phys. Rev. B} {\bf 77}, 125424 (2008).
\bibitem {Catalan4} H. Lu, C. Bark, D. de los Ojos, J. Alcala, C. Eom, G. Catalan, and A. Gruverman, Mechanical writing of ferroelectric polarization, {\it Science} {\bf 336}, 59 (2012).
\bibitem {martin} R. M. Martin, Piezoelectricity, {\it Phys. Rev. B} {\bf 5}, 1607 (1972).
\bibitem {Tagantsev5} A. K. Tagantsev, Theory of  flexoelectric effect in crystals, {\it Sov. Phys. JETP} {\bf 61}, 1246 (1985).
\bibitem {Resta1} R. Resta, Towards a Bulk Theory of Flexoelectricity, {\it Phys. Rev. Lett.} {\bf 105}, 127601 (2010).
\bibitem {Tagantsev} A. K. Tagantsev, Piezoelectricity and  flexoelectricity in crystalline dielectrics, {\it Phys. Rev. B} {\bf 34}, 5883 (1986).
\bibitem {Sharma1} R. Maranganti and P. Sharma, Atomistic determination of flexoelectric properties of crystalline dielectrics, {\it Phys. Rev. B} {\bf 80}, 054109 (2009).
\bibitem {Cross6} W. Ma and L. E. Cross, Flexoelectric effect in ceramic lead zirconate titanate, {\it Appl. Phys. Lett.} {\bf 86}, 072905 (2005).
\bibitem {vasp} G. Kresse and J. Furthmuller, Efficient iterative schemes for ab initio total-energy calculations using a plane-wave basis set, {\it Phys. Rev. B} {\bf 54}, 11169 (1996).
\bibitem {qe} P. Giannozzi, S. Baroni, N. Bonini, M. Calandra, R. Car, C. Cavazzoni, D. Ceresoli, G. L. Chiarotti, M. Cococcioni, I. Dabo, A. Dal Corso, S. de Gironcoli, S. Fabris, G. Fratesi, R. Gebauer, U. Gerstmann, C. Gougoussis, A. Kokalj, M. Lazzeri, L. Martin-Samos, N. Marzari, F. Mauri, R. Mazzarello, S. Paolini, A. Pasquarello, L. Paulatto, C. Sbraccia, S. Scandolo, G. Sclauzero, A. P. Seitsonen, A. Smogunov, P. Umari, and R. M. Wentzcovitch, Quantum espresso: a modular and open-source software project for quantum simulations of materials, {\it J. Phys. Condens. Matter} {\bf 21}, 395502 (2009).
\bibitem {gbt1} D.-B. Zhang and S.-H. Wei, Inhomogeneous strain induced half-metallicity in bent zigzag graphene nanoribbons, {\it Npj Comput. Mater.} {\bf 3}, 32 (2017).
\bibitem {gbt2} L. Yue, G. Seifert, K. Chang, and D.-B. Zhang, Effective zeeman splitting in bent lateral heterojunctions of graphene and hexagonal boron nitride: A new mechanism towards half-metallicity, {\it Phys. Rev. B} {\bf 96}, 201403 (2017).
\bibitem {gbt3} D.-B. Zhang, G. Seifert, and K. Chang, Strain-induced pseudomagnetic fields in twisted graphene nanoribbons, {\it Phys. Rev. Lett.} {\bf 112}, 096805 (2014).
\bibitem {film1} Y. Umeno, A. Kushima, T. Kitamura, P. Gumbsch, and J. Li, Ab initio study of the surface properties and ideal strength of (100) silicon thin films, {\it Phys. Rev. B} {\bf 72}, 165431 (2005).
\bibitem {film2} M. Milosavljevic, C. Jeynes, and I. H. Wilson, Epitaxial (100) silicon films grown at low temperatures in an electron-beam evaporator, {\it J. Appl. Phys.} {\bf 57}, 1252 (1985).
\bibitem {dftb1} D. Porezag, T. Frauenheim, T. Kohler, G. Seifert, and R. Kaschner, Construction of tight-binding-like potentials on the basis of density-functional theory: Application to carbon, {\it Phys. Rev. B} {\bf 51}, 12947 (1995).
\bibitem {dftb2} R. Rurali and E. Hernandez, Trocadero: a multiplealgorithm multiple-model atomistic simulation program, {\it Comput. Mater. Sci.} {\bf 28}, 85 (2003).
\bibitem {electric} M. Grundmann, The physics of semiconductors, {\it (Springer-Verlag, 2006).}
\bibitem {wei1} Y.-H. Li, X. G. Gong, and S.-H. Wei, Ab initio allelectron calculation of absolute volume deformation potentials of iv-iv, iii-v, and ii-vi semiconductors: The chemical trends, {\it Phys. Rev. B} {\bf 73}, 245206 (2006).
\bibitem {wei2} S.-H. Wei and A. Zunger, Role of metal d states in ii-ui semiconductors, {\it Phys. Rev. B} {\bf 37}, 8958 (1988).
\bibitem {Resta4} R. Resta, Deformation-potential theorem in metals and in dielectrics, {\it Phys. Rev. B} {\bf 44}, 11035 (1991).
\bibitem {Resta3} R. Resta, Macroscopic polarization in crystalline dielectrics: the geometric phase approach, {\it Rev. Mod. Phys.} {\bf 66}, 899 (1994).
\bibitem {Resta2} R. Resta, Electrical polarization and orbital magnetization: the modern theories, {\it J. Phys. Condens. Matter} {\bf 22}, 123201 (2010).
\bibitem {polar1} N. A. Spaldin, A beginner's guide to the modern theory of polarization, {\it J. Solid State Chem.} {\bf 195}, 2 (2012).
\bibitem {Tagantsev2} A. K. Tagantsev and A. S. Yurkov, Flexoelectric effect in finite samples, {\it J. Appl. Phys.} {\bf 112}, 044103 (2012).
\bibitem {Stengel1} M. Stengel, Microscopic response to inhomogeneous deformations in curvilinear coordinates, {\it Nature Commun.} {\bf 4}, 2693 (2013).
\bibitem {surface} X. Zhang, Q. Pan, D. Tian, W. Zhou, P. Chen, H. Zhang, and B. Chu, Large Flexoelectric like Response from the Spontaneously Polarized Surfaces in Ferroelectric Ceramics, {\it Phys. Rev. Lett.} {\bf 121}, 057602 (2018).
\bibitem {Cross4} W. Ma and L. E. Cross, Flexoelectricity of barium titanate, {\it Appl. Phys. Lett.} {\bf 88}, 232902 (2006).
\bibitem {Cross7} W. Ma and L. E. Cross, Strain-gradient-induced electric polarization in lead zirconate titanate ceramics, {\it Appl. Phys. Lett.} {\bf 82}, 3293 (2003).
    
\end{thebibliography}
\end{document}